%% file: PASJ-WB724_060830.tex
\documentclass{pasj00}

\def\kms{~km~s$^{-1}$}
\def\h2o{H$_2$O}
\def\vlsr{$V_{\mbox{\scriptsize LSR}}$\ }
\def\etal{ et~al.\ }
\def\g192{G192.16$-$3.84}
\def\j0603{J0603+1742}

\draft
\SetRunningHead{H.~Imai\etal}{Water Masers in \g192}

\title{A collimated jet and an infalling-rotating disk in \g192 traced by 
\h2o  maser emission}
\author{Hiroshi \textsc{Imai}\altaffilmark{1},
Toshihiro \textsc{Omodaka}\altaffilmark{1}, 
Tomoya \textsc{Hirota}\altaffilmark{2}, 
Tomofumi \textsc{Umemoto}\altaffilmark{3}, \\
Kazuo \textsc{Sorai}\altaffilmark{4}, and 
Tetsuro \textsc{Kondo}\altaffilmark{5}}
\altaffiltext{1}{Department of Physics, Faculty of Science, 
Kagoshima University, Kagoshima 890-0065}
\altaffiltext{2}{Mizusawa VERA Observatory Mitaka Office, National Astronomical 
Observatory, Mitaka, Tokyo 181--8588}
\altaffiltext{3}{Nobeyama Radio Observatory, 
National Astronomical Observatory,\\
Minamimaki, Minamisaku, Nagono 384-1305}
\altaffiltext{4}{Division of Physics, Graduate School of Science, 
Hokkaido University, Sapporo 060-0810}
\altaffiltext{5}{Kashima Space Research Center, 
National Institute of Information and Communication Technology, \\
Kashima, Ibaraki 314-0012}
\email{(HI) hiroimai@sci.kagoshima-u.ac.jp}
\Received{2006/5/18}
\Accepted{2006/8/14}
\Published{2006/10/25}

\begin{document}
\setlength{\baselineskip}{6ex}

\maketitle

\begin{abstract}
We report \h2o masers associated with the massive-star forming region \g192  
observed with the new Japan VLBI network at three epochs spanned for two months, 
which have revealed the three-dimensional kinematical structure of the whole \h2o  maser region in \g192, containing two young stellar objects separated by 
$\sim$1200~AU. The maser spatio-kinematical structure has well persisted since previous observations, in which the masers are expected to be associated 
with a highly-collimated bipolar jet and an infalling-rotating disk in the northern 
and southern clusters of \h2o maser features, respectively. 
We estimated a jet expansion speed of $\sim$100\kms\ and re-estimated a 
dynamical age of the whole jet to be 5.6$\times$10$^{4}$~yrs. 
We have investigated the spatial distribution of Doppler velocities during the 
previous and present observations and relative proper motions of \h2o\ 
maser features in the southern cluster, and a relative bulk motion between the 
two maser clusters. They are well explained by a model of an infalling--rotating 
disk with a radius of $\sim$1000~AU and a central stellar mass of 5--10~$M_{\odot}$, 
rather than by a model of a bipolar jet perpendicular to the observed CO outflow. 
Based on the derived \h2o maser spatio-kinematical parameters, we discuss the 
formation mechanism of the massive young stellar objects and the outflow development in \g192. 
\end{abstract} 

\section{Introduction}
\label{sec:introduction}
Formation mechanism of massive stars is one of the most important issues in astrophysics for elucidating their whole evolution as well as their effects on dynamical and chemical composition evolution of galaxies and the universe. It is 
difficult, however, to observationally find the true mechanism because all of massive 
young stellar objects (MYSOs) are associated with star clusters that are deeply embedded in molecular clouds located at large distances from the Sun ($d\gtrsim$500~pc). Therefore, in whichever scenario of massive star formation, observations with very high angular resolution (better than 10~milliarcseconds, mas) is essential to resolve a massive-star forming region into individual MYSOs. Because a massive star also 
evolves extremely rapidly, it is difficult to find MYSOs at the earliest stage 
of the star formation process. An MYSO with $M\gtrsim 8\:M_{\odot}$ has a 
Kelvin-Hermholtz time scale that is shorter than a free-fall time scale, therefore 
it reaches a zero-age main sequence star while mass accretion still occurs. 
There exists a basic question how such an MYSO builds its mass against mass 
ejection by strong stellar wind and expansion of an H{\rm II} region. 

There are two major possible scenarios of massive star formation; the first 
in term of a massive and thick accretion gas disk/torus and the second in term of lower-mass stars merged into a higher-mass star. In the first scenario, a parent molecular cloud core is still collapsing and providing material into an MYSO 
with a very high mass-accretion rate (e.g. $\dot{M}\gtrsim 10^{-3}\: 
M_{\odot}$~yr$^{-3}$) by way of a thick gas disk even if the central part of the 
core is photoionized (e.g., \cite{ket03}). The presence of a disk near a massive star 
can significantly reduce the effect of radiation on the infalling material by reducing 
the exposure of matter to the radiation field and by allowing photons to escape along the stellar polar axis (e.g. \cite{nag89,kru05}). A direct evidence for this scenario should be to directly detect mass accretion onto the surface or in very vicinity of an MYSO within $\sim$1~AU. Previous observations have reported the existence of such massive disks/tori (e.g. \cite{sak05,jia05,ket06}), but their physical scales 
directly elucidated by the observations are still much larger ($>$100~AU). Nevertheless these observations suggested that a single star with a mass at least 
equal to or lower than 7~$M_{\odot}$ can be created in this scenario \citep{jia05}. The formation mechanism of a higher-mass star is still obscure. \citet{chi04} reported an MYSO in M17 with a stellar mass of 20~$M_{\odot}$ and an accompanying massive gas/dust disk, but the estimated mass is reevaluated by \citet{sak05} 
to be only 3--8~$M_{\odot}$. In the second scenario, the "runaway collapse" is 
expected, in which two stars are merged while the 
third companion is kicked away so that an angular momentum is released by the 
companion from this triple star system (e.g., \cite{bon04,bon06}). A massive 
star is formed as a result of a series of events in which two lower-mass stars 
are merged. This scenario has been also indirectly supported by observational 
results, in which an MYSO and other young stellar objects (YSOs) are moving from 
almost the same position on the sky (in Orion KL/BN region, \cite{rod05,gom05}). 
An explosion may have happened when these stellar objects were located in the 
common point with strong dynamical interactions \citep{bal05}. However, the 
stellar volume density requested for this scenario ($>$10$^{7}$~pc$^{-3}$) 
should predict a high frequency of such runaway collapse and explosion events. 
More observational examples are necessary to widely support this scenario as a 
major scenario of massive stars. 

\h2o maser emission is one of the most important probes for study on star 
formation, often based on data obtained by using very long baseline 
interferometry (VLBI) with high angular and velocity resolution. Analyses of 
spatial positions, Doppler velocities, and proper motions of individual maser 
features with a typical size of 1~AU \citep{rei81} have revealed the 3-D gas 
kinematics often in very vicinity ($<<$100~AU) of YSOs  
(e.g.\ \cite{cla98,fur00,set02, tor01, tor03, mos05, fur05, god04, god06}).
In practice, details of the gas kinematics are complicated but seem to depend on 
mainly evolutionary status of YSOs. \h2o maser sources sometimes enable us to 
measure internal motions of giant molecular clouds by measuring relative bulk motions 
between clusters of \h2o  masers that are separated by up to 1~pc but still located 
within a single antenna beam (e.g., \cite{ima00,tor01}). Such bulk motions may 
be owing to, e.g.,  propagation of shock layers from newly-formed H{\rm II} regions, cloud contraction by the self-gravitation due to a huge mass of a giant molecular 
cloud, or runaway motions as mentioned above. 

The massive-star forming region \g192 (hereafter G192) is located at a distance 
of $\sim$2~kpc, and has massive YSO candidates, one of which is a B2-type star with 
a giant Herbig-Haro bipolar outflow \citep{dev99} and a possible circumstellar disk 
(\cite{she99}, hereafter SK99; \cite{she01}, hereafter SCK01; \cite{ind03}; 
\cite{she04}, hereafter SBCSK04). There are two clusters of \h2o masers separated by $\sim$0.\arcsec6 ($\sim$1200~AU) 
in the north--south direction (SK99; SBCSK04). The northern and southern clusters 
exhibit alignments of maser features parallel and perpendicular to two outflows, 
which are traced by $^{12}$CO emission that are projected roughly in the east--west direction parallel to the Herbig-Haro objects mentioned above (\cite{she98}, hereafter SWSC98). On the basis of the spatial distribution of \h2o\ masers and C$^{18}$O emission tracing high density gas, SK99 and SBCSK04 have proposed that the southern \h2o\ maser features are associated with a flattened rotating gas torus around a B2 star, in which the maser velocities are roughly consistent with those in a Keplerian disk. 

The Very Long Baseline Array (VLBA) observations of the G192 
\h2o masers were also reported by SBCSK04, but a maser proper motion measurement, 
especially for the southern cluster of \h2o masers, was impossible because 
quenching the maser activity. 

Here, we report short-term monitoring observations towards the G192 \h2o   
masers with new Japan VLBI network (JVN, Fujisawa \etal, in preparation), 
which is newly organized and now consisting of four 20~m telescopes of the 
VLBI Exploration of Radio Astrometry (VERA), a 45~m telescope of Nobeyama 
Radio Observatory\footnote{NRO and VERA observatory are branches of the National 
Astronomical Observatory, an interuniversity research institute operated 
by the Ministry of Education, Culture, Sports, Science and Technology.}, 
a 34~m telescope of Kashima Space Research Center (KSRC), and other 11--64~m 
dish telescopes in Japan. Fortunately we could 
detect maser proper motions for both of the maser clusters, leading us to 
obtaining more direct evidence for the circumstellar disk. Sect.\ 2 describes 
the VLBI observations and data reduction. Sect.\ 3 summarizes analyses of the 
revealed 3-D kinematics of the \h2o\ masers. Sect.\ 4 discusses the physical 
conditions and evolutionary statuses of the jet and disk objects in G192. 
  

\section{Observations and Data Reduction}
\label{sec:observation}
The VLBI observations were made at three epochs during 2005 March--June, 
using five or six telescopes of the JVN. Table \ref{tab:status} gives the 
status of these observations. In each epoch, the observation was made for 
8--10 hours including scans of G192 and calibrators (J0530$+$1331 
and DA~193). The received signals were recorded with the SONY DIR1000 recorder 
with a rate of 128M bits~s$^{-1}$ and in two base band channels with a band 
width of 16 MHz each. The VERA telescopes simultaneously observed not only the 
sources mentioned above but also the position reference source \j0603 
with the dual beam system (e.g. \cite{hon03}). The individual base band channels 
were assigned to observations of either the maser or reference source. 

The data correlation was made with the Mitaka FX correlator \citep{chi91}. 
An accumulation period of the correlation was set to 1~s. The correlation 
outputs consisted of 1024 velocity channels, each of which has a velocity 
spacing of 0.21\kms. From the VERA telescopes, there were two correlation 
outputs including scans of G192 and the calibrators in the first and 
the reference source in the second, respectively. 

Data reduction to obtain image cubes for the maser source was made with the 
NRAO AIPS package in normal procedures (e.g., \cite{dia95}). Fringe fitting 
and self-calibration procedures were made for a Doppler velocity channel 
including a bright maser spot (velocity component) in G192, which is also given 
in Table \ref{tab:status}. The solutions were applied to all of the data in 
other velocity channels. A typical size of the synthesized beam was 1~milliarcseconds 
(mas) in the three observations (see Table \ref{tab:status}). A relative position 
accuracy of a maser spot was ranged in 0.01--0.09 mas depending on a 
signal-to-noise ratio and spatial structure of the spot. Identification of an 
\h2o  maser feature (a cluster of maser spots or velocity components) was made 
in the same procedure shown in several previous papers (e.g. \cite{ima00}), in which 
the feature position is defined as a brightness peak in the feature. A relative 
position accuracy of a maser feature was ranged in 0.03--0.17 mas. Relative proper 
motions were measured for maser features identified at least at two of the 
three epochs. 

The position reference source \j0603 was marginally but reliably detected at only the first epoch with the {\it inverse phase-referencing technique}, in which 
fluctuation of visibility phases in the \j0603 data were compensated by those in the 
spectral channel (see the phase-reference velocity channel shown in table 
\ref{tab:status}) that included the brightest maser emission in the feature 
\g192:I2006 {\it 9}. \j0603 had a correlated peak intensity of 
50~mJy~beam$^{-1}$ with a beam tapered by a $uv$-distance of 80~M$\lambda$.

\section{Results}


\subsection{Summary of the revealed maser spatio-kinematical structure}
\label{sec:results-summary}

\begin{figure}
  \begin{center}
    \FigureFile(155mm, 200mm){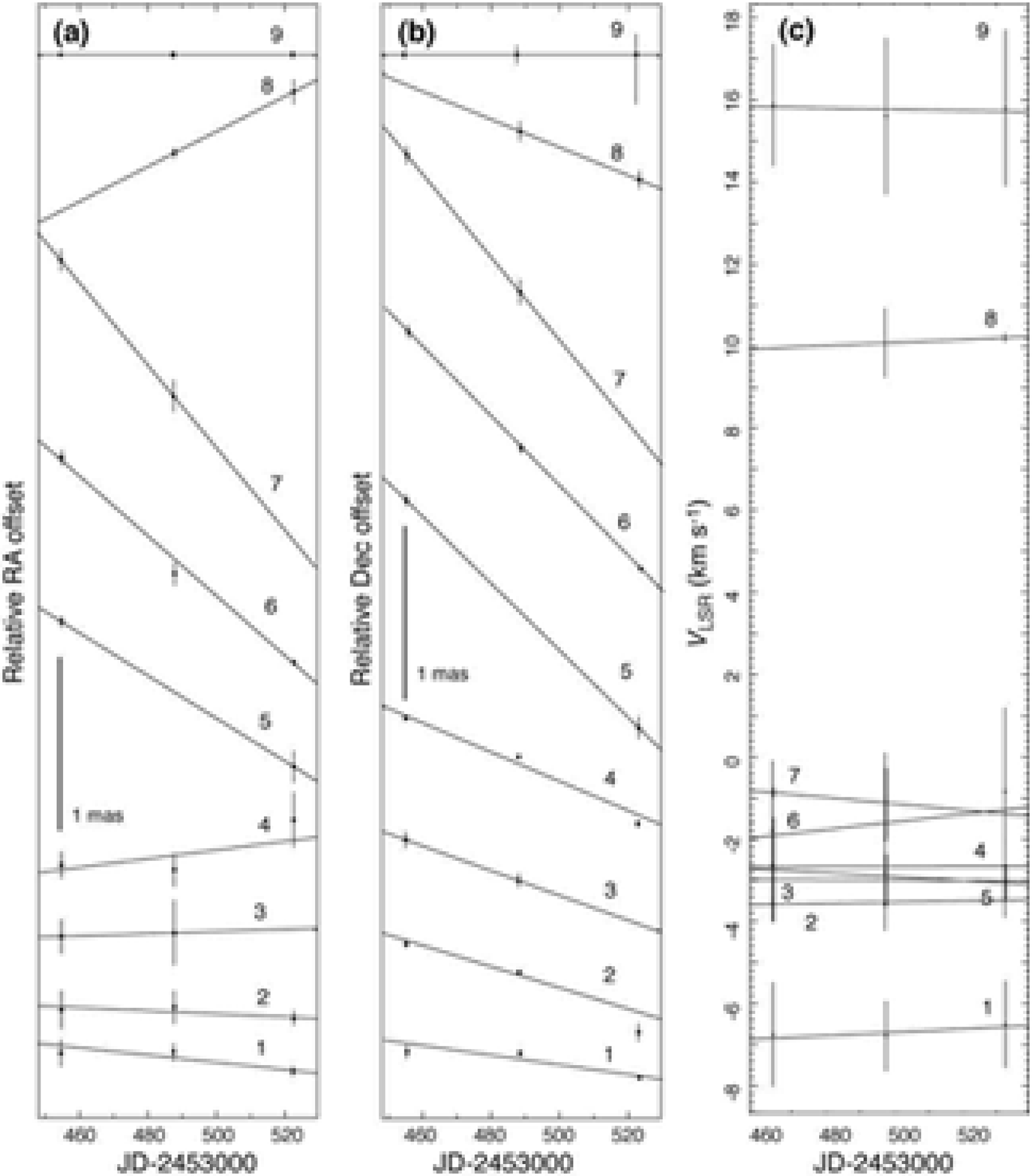}
  \end{center}
\caption{Observed relative proper motions (sub-panels a and b) and Doppler velocity 
variations (sub-panel c) of \h2o  maser features in \g192. A number added for each 
feature in each sub-panel shows the assigned one after the designated name form ``\g192: I2006" A solid line in the position or Doppler velocity plot indicates a least-square-fitted line assuming a constant velocity motion in the proper motion or a constant Doppler velocity drifts.}
\label{fig:PM-WB724}
\end{figure}

Table \ref{tab:status} shows the numbers of the detected \h2o maser features in 
the G192 region. Although the \h2o maser emission in G192 is highly time-variable, 
we could detect most of velocity components found in the previous VLA/VLBA 
observations (SK99; SBCSK04). {\it Relative} proper motions of eight maser 
features with respect to the maser feature \g192:I2006 {\it 1} (see table \ref
{tab:pmotions}) as position reference were identified. Table \ref{tab:pmotions} 
lists the measured proper motions. Figure \ref{fig:PM-WB724} shows time 
variations of positions and Doppler velocities of the maser features that are 
detected their relative proper motions. 

\begin{figure*}
  \begin{center}
    \FigureFile(85mm, 110mm){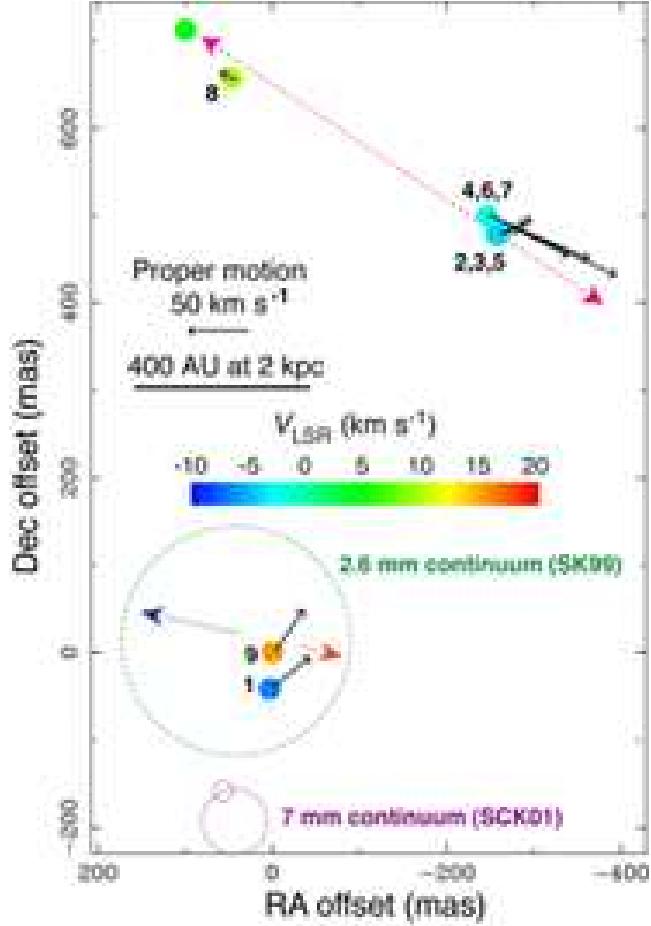}
  \end{center}
  \caption{Doppler velocity distribution (colored filled circle) and relative 
proper motions (black arrow) of \h2o  masers in \g192. The displayed proper motion vector is that subtracted by a velocity bias ($\overline{\Delta V_{x}}, 
\overline{\Delta V_{y}})=(23,-34)$[\kms] from the original vector so that the 
northern jet source is spatially fixed. A number added for each feature 
with a proper motion shows the assigned one after the designated name form ``\g192: I2006". The map origin is located at 
R.A.(J2000.0)$=$05$^{h}$58$^{m}$13$^{s}$\hspace{-2pt}.5332, 
decl.(J2000.0)$=+$16\arcdeg31\arcmin58\arcsec.483,  
which was estimated as described in the main text. The locations of 2.6~mm and 7~mm continuum emission peaks estimated by SK99 and SCK01, respectively, are roughly 
indicated by dashed circles. Axes of the northern jet and the outer outflow proposed 
by SBCSK04 are indicated by dashed arrow.} 
\label{fig:WB724-color}
\end{figure*}

We found a flow such as a highly-collimated bipolar jet in the maser 
spatio-kinematical structure in the northern maser cluster. We estimated a bias of 
the proper motion vectors with respect to the intrinsic motion of the position-reference feature,  
$(\bar{V_{x}}, \bar{V_{y}})=$(23,$-$34)[km~s$^{-1}$], by calculating a mean of 
two velocity vectors that are equally-weighted average proper motions of the NE 
and SW clusters of maser features, respectively, in the northern cluster. Figure 
\ref{fig:WB724-color} shows the distribution and the relative proper motion 
fields of \h2o masers in G192, in which the proper motion vectors are subtracted 
by the velocity bias so that a systemic motion of the northern jet is removed. 
A Doppler velocity bias, likely corresponding to the systemic radial velocity 
of the source driving the maser jet is about \vlsr$=$~5\kms, which is 
consistent with the systemic velocity of the CO molecular outflow seen on a much 
larger scale in roughly the same direction within 1\kms (e.g., \cite{sne90}; SK99). 

Fortunately, we also detected two maser features $\sim$1200~AU away from 
the bipolar jet with Doppler velocities \vlsr$=$ $-$6.8\kms and 16.0\kms 
(\g192:I2006 {\it 1} and {\it 9}, hereafter abbreviated as the features 
A and B, respectively). The both features are moving toward the bipolar jet, 
but the feature B is relatively receding by $\sim$10\kms\ in the SSW direction 
with respect to the feature A. The positions and the relative proper 
motion of the two features provide strong constraints on kinematical models 
of the masers discussed in Sect. \ref{sec:disk-model}, \ref{sec:rmotion}. 


\subsection{Astrometry for the \h2o maser source}
\label{sec:astrometry}

As described in Sect.\ref{sec:observation}, we successfully detected the 
position reference source \j0603 to obtain the absolute coordinates of the 
\g192 \h2o masers. Based on the measured coordinate offsets of \j0603 with 
respect to its true coordinates, we estimated coordinates of the maser 
feature B to be, R.A.(J2000.0)$=$05$^{h}$58$^{m}$13$^{s}$\hspace{-2pt}.5332, 
decl.(J2000.0)$=+$16\arcdeg31\arcmin58\arcsec.483. Taking into account 
difference in the coordinates obtained by some data analyses adopting 
different analyzing parameters, an uncertainty of the measured coordinates 
is expected to be about 6~mas. This accuracy is still much worse than those 
achieved by the latest results of VERA astrometry (Honma  \etal, Hirota 
\etal, and Imai \etal in 2006, in paper preparation), but does not affect 
any interpretation in the present paper. 

This astrometry enables us to superpose the \h2o maser maps made from previous 
VLA/VLBA and our JVN observations with an accuracy better than 100~mas. Our 
\h2o maser maps can be superposed on the 7~mm continuum map with an accuracy 
of about 20~mas. These accuracies are estimated from the astrometric accuracies 
of previous VLA observations (SCK01; SBCSK04) and uncertainty of a secular motion 
of the maser source due to the Galactic rotation and an unknown local motion. 
The maser features A and B were located closest to the positions of the maser 
features detected by SK99 at a Doppler velocity \vlsr$\simeq$10\kms. The 
superposed distribution of the \h2o  masers together with relative proper 
motions reveals more clearly the kinematics of the \h2o  masers in the 
northern and southern clusters. 

\subsection{Objective analyses of the jet kinematics for the northern 
maser cluster}
\label{sec:vvcm} 

In order to objectively or model-independently find an axis of a jet  
accompanying the northern maser cluster, we performed diagonalization for 
the velocity variance--covariance matrix (VVCM) obtained from velocity 
vectors of maser features (c.f., \cite{blo00}). The eigenvector corresponding 
to the largest eigenvalue (velocity dispersion) indicates a major axis of 
the jet. A square root of the largest and second largest eigenvector gives 
a collimation factor of the jet. Uncertainties of the obtained eigenvectors 
and eigenvalues were derived from standard deviations of these parameters, 
which were calculated by the Monte--Carlo simulation for the VVCM 
diagonalization using velocity vectors randomly distributed around the 
observed values within their estimated errors. Table \ref{tab:vvcm} gives the 
obtained eigenvalues and their corresponding eigenvectors after the matrix 
diagonalization. In the cases of G192, the eigenvector corresponding to the 
largest eigenvalue has an inclination of 2\arcdeg\ with respect to the sky plane 
and a position angle of 60\arcdeg\ east from the north. This vector is almost parallel 
to outflows found in the Herbig-Haro objects and $^{12}$CO emission on a much larger 
scale. Proper motions with a systemic velocity bias subtracted (see figure 
\ref{fig:WB724-color}) are also roughly parallel to this vector. These suggest that 
the maser jet and the CO outflow should be in the same outflow without significant precession.

Similar to the VVCM, the spatial variance--covariance matrix is obtained from 
the maser feature distribution, which also gives an eigenvector direction similar 
to that of the major axis for the VVCM. Taking into account a velocity 
collimation factor of 4.9 and the bipolarity of the velocity vectors, the 
maser spatio-kinematical structure indicates a highly-collimated bipolar jet. 
Note that maser velocity vectors in the south-west cluster in the bipolar jet 
are directed toward both directions of the bipolar jet. This maser cluster also 
contains both of the red-shifted and the blue-shifted maser components in the maps 
obtained by SBKSC04. Note that the large scale molecular outflows are estimated to 
be driven by the southern YSO having the southern maser cluster, not by the northern YSO having this northern maser cluster (SWSC98; \cite{dev99}). The VVCM analysis for the \h2o\ masers suggests that there are two outflows in \g192 that are parallel, one of which is highly collimated and almost completely parallel to the maser jet. 

\subsection{Dynamical ages of the bipolar outflows in \g192}
\label{sec:vvcm} 
We estimated dynamical ages of the bipolar outflows driven by the northern 
and southern YSOs in G192 on basis of the 3-D kinematical structure of the 
\h2o masers that are associated with the outflow driven by the northern object 
(see sect.\ \ref{sec:results-summary}). The outflow found 
in the Herbig-Haro objects and $^{12}$CO emission is likely to be driven by 
the southern object (SWSC98) and the outflow driven by the northern object 
may not directly seen on the same scale. Nevertheless the latter is expected 
to be included in the $^{12}$CO emission and to have a length equal to or 
shorter than that of the former. The $^{12}$CO emission flow also extends 
in the north--south directions by $\sim$0.6~pc and seems to have two axes of 
outflows with similar projected lengths. One of the outflow axes might trace 
the outflow driven by the northern YSO. If the northern and southern objects 
started their star formation process from a common molecular cloud core, 
these outflows may simultaneously ignite and have the same dynamical age, 
and eventually roughly the same outflow velocity. 

The length of the southern outflow projected on the sky is estimated to be 
$\sim$5.7~pc from the existence of the Herbig-Haro objects HH396/397 (\cite{dev99}, see also the $^{12}$CO emission map of SWSC98). This length is an upper limit of that of the northern outflow. On the other hand, the velocity of the northern outflow is estimated to be $\sim$100\kms\ in the present work from the blue-shifted maser feature \g192:I2006 {\it 7} that is moving fastest from the vicinity of the flow's kinematical center (see figure \ref{fig:WB724-color}). Therefore, assuming a constant velocity in the whole outflow, a dynamical age of the outflows is about $5.6\times10^{4}$~yrs. 
This value is shorter by a factor of 3.5 than those previously estimated without 
any information of the 3-D flow velocity ($\sim 2 \times 10^{5}$~yrs, \cite{sne90}; SWSC98; \cite{dev99}). 

\subsection{A relative bulk motion between the northern and southern maser 
clusters in \g192}
\label{sec:rmotion} 

In the present work, we clearly detected a relative bulk motion, in which the 
northern and southern clusters of \h2o\  masers are approaching each other by 
$\sim$40\kms. Figure \ref{fig:WB724-color} shows that in this velocity the 
features A and B are approaching the northern maser cluster whose systemic 
motion is removed in this figure (see sect.\ \ref{sec:results-summary}). If the 
YSO located at the dynamical center of the northern outflow is close to the south-west cluster of maser cluster as mentioned in sect.\ \ref
{sec:results-summary}, the direction of the bulk motion is exactly along the alignment of the two YSOs. Thus it is possible that these two YSOs might be dynamically 
linked. However the approaching speed is extraordinary large and it is unlikely that 
this bulk motion is an orbiting motion between the two YSO. When supposing this motion 
as an orbiting motion, a possible lower limit of an enclose mass in this system 
should be as large as 1700~$M_{\odot}$, such a huge mass within such a small 
scale ($\sim$1200~AU) has never been observed in this region. On the other hand, 
the systemic velocities of the two YSO may be almost equal within 1\kms\ as mentioned 
in sect.\ \ref{sec:results-summary} and \ref{sec:disk-model}. Therefore, a true 
relative bulk motion between the two YSOs should be much smaller, about 2--3\kms, 
which corresponds to an enclose mass of about 10~$M_{\odot}$ in this system. 

The possible existence of an outflow parallel to the relative maser bulk 
motion may also be rejected. The \h2o maser features associated with the 
southern YSO and detected by previous and present works are aligned over 
1000~AU. in the north--south direction. The alignment is also parallel to 
that found in C$^{18}$O emission on an angular scale of 10000~AU (SK99). 
The dynamical center of the masers should be one of the continuum emission 
sources detected in 3.6~cm, 1.3~cm, 7~mm, and 2.6~mm bands, which are 
located within $\sim$400~AU from the masers. It is impossible to form a 
straight alignment of masers longer than this scale by a shock wave front 
created by an outflow driven perpendicular to the alignment. A major axis 
of a (highly-collimated) outflow might be exist along this alignment. However, 
the distributions of Herbig-Haro objects and $^{12}$CO emission indicate the 
existence of outflows only in the roughly east--west direction (SWSC98; 
\cite{dev99}). It is unreasonable to interpret that only C$^{18}$O emission 
is detected in the north--south direction as a tracer of a high density outflow. 

$^{13}$CO emission mapped by SWSC98 may trace expanding shells created along 
the interface between the outflow and ambient cloud gas. 
The opening angle of the shells from the axis is 45\arcdeg\ and one of 
the shells is aligned roughly parallel to the direction of the maser bulk 
motion. However, the observed bulk motion is much faster than the shell 
expansion ($\sim$8\kms). Except for the possibility of a binary system in the 
southern YSO itself, it is most feasible to interpret that the measured 
relative maser bulk motion should be coming from another internal motions of 
the maser features, such as motions expected from a rotating disk model 
discussed in the next section.

\subsection{An infalling-rotating thin disk model for the southern YSO}
\label{sec:disk-model}

The previous VLBA and VLA observations (SK99; SBCSK04) have detected more 
than 10 \h2o\ maser features in the southern cluster of features and found 
the maser alignment in the north--south direction, which is roughly 
parallel to those of the elongated morphology and a velocity gradient seen 
in C$^{18}$O emission (SK99). SK99 found a clearer 
velocity gradient in the maser distribution and fitted it to a model of a 
Keplerian rotating gas disk whose center is located around the peak of 
7~mm/3.6~cm continuum emission 
in \g192, or likely a B-type YSO. Later SBCSK04 found 
maser components with significant red shift (11$\geq$\vlsr$\geq$15\kms) in the 
east side of the maser alignment found by SK99, which is inconsistent with 
the locations expected from a bipolar outflow in the east--west direction 
with an opposite velocity gradient (SWSC98) or another possible outflow 
parallel to the maser alignment. Such a complicated maser distribution 
may be explained by combination of outflow, infall, and rotation motions. 

Here we attempt to obtain a spatio-kinematical model for the \h2o masers 
detected in the previous and present observations. We adopt a disk 
model proposed by \citet{cas81} and \citet{fie97}, in which gas clumps are 
moving in the trajectory planes with a specific angle with respect to the 
axis perpendicular to the disk and impinging onto the disk. \h2o maser 
emission is expected to be excited on the disk where an infalling gas clump 
collides with the disk and reaches an appropriate physical condition for 
maser excitation 
($T\simeq$400~K and $n_{\mbox{H$_{2}$}}\simeq10^{9}$~cm$^{-3}$) by shock.  
The velocity field of the disk $(v_{r}, v_{\theta}, v_{\phi})$ is calculated 
using the zero-energy orbits of the impinging clumps on parabolas, and is 
expressed in the spherical coordinate system as follows, 

\begin{eqnarray}
v_{r} & = & -\left(\frac{GM_{\ast}}{r} \right)^{1/2} \\
v_{\theta} & = & \left(\frac{GM_{\ast}}{r} \right)^{1/2}\cos\theta_{0} \\
v_{\phi} & = & \left(\frac{GM_{\ast}}{r} \right)^{1/2}\sin\theta_{0} \\
\sin^{2}\theta_{0} & = & r/r_{\mbox{d}}
\label{eq:model3}.
\end{eqnarray} 

\noindent
Here $r$ is the distance from the central object on the disk, $M_{\ast}$ the 
mass of the central stellar object, $\theta_{0}$ the angle between the disk rotation 
axis and the trajectory plane of the clump, and $r_{\mbox{d}}$ the distance from 
the stellar object where $\theta_{0}=0$. Note that the velocity component 
$v_{\theta}$ should be zero for the clump on the disk with maser emission 
excited after collision with the disk. To examine the disk model proposed 
previously, we adopt physical parameter values that are similar to those 
estimated previously. A systemic velocity of 
\vlsr$=$5.7\kms\ assumed here is estimated from the C$^{18}$O  and NH$_{3}$ 
emission detected with radio interferometers (SK99; SBCSK04). A disk 
inclination with respect to the line of sight is estimated to be 
$\theta\simeq$63\arcdeg (SWSC98; SK99). An outer disk 
radius $r_{\mbox{d}}$ may be equal to or smaller than 1000~AU 
(0\arcsec.5 at 2~kpc), this upper limit corresponds to a separation between 
the northern and southern clusters of the \h2o  masers. 

\begin{figure*}
  \begin{center}
    \FigureFile(160mm, 90mm){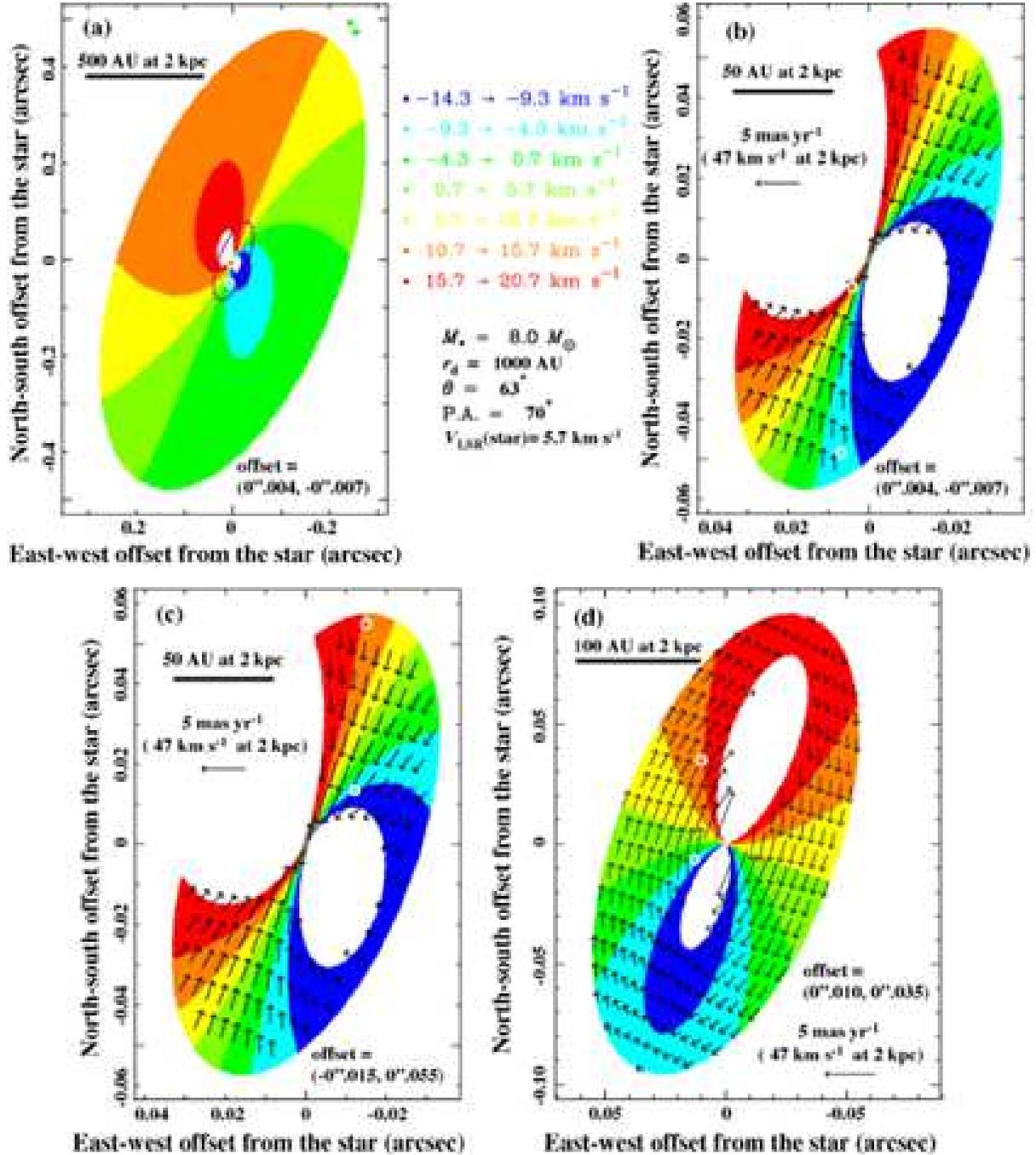}
  \end{center}
  \caption{Calculated LSR-velocity distribution for masing gas clumps on 
the modeled thin disks. The \h2o  maser features displayed in figure 
\ref{fig:WB724-color} (filled circles) are overlaied after shifting the offset 
coordinates by the value shown in the sub-panels. All of the models adopt the same physical parameters of the disk, except for the velocity field. (a) A disk model with 
impinging maser clumps, which is best consistent with the observed maser kinematics.  The two maser features are located at the southern part on the disk. The whole disk with an apparent radius of $\leq$0\arcsec.5 is displayed. A dashed purple ellipse 
indicates a displayed region of the same disk in (b). (b) Same as (a) but displaying 
the inner part of the disk. An arrow indicates a proper motion of the gas clump located 
at the root of the arrow. (c) Same as (b) but the maser features are located at 
the northern part on the disk. (d) A purely Keplerian-rotating disk model.}
\label{fig:model}
\end{figure*}

Through several trials of the model fitting, we have found that a position angle 
of the disk axis, P.A.$=$60--75\arcdeg\ east from the north, and a stellar mass of 
5--10~$M_{\odot}$ provides a model consistent with the observed 
3-D maser kinematical structure. These two parameters estimated are consistent 
with those previously suggested ($\sim$75\arcdeg and  6--10~$M_{\odot}$, 
respectively, SK99; SCK01). Figure \ref{fig:model} shows the spatial distribution 
of maser Doppler velocities which are expected in one of the best consistent model. 
Although figure \ref{fig:model} shows the distribution of only the \h2o\ 
masers detected by the present observations, the proposed model 
(figure \ref{fig:model}{\it a} and {\it b}) may  well explain the Doppler 
velocity distribution of the maser features that have been detected both by 
previous and present works. The modeled radial velocity field is also 
roughly consistent with that found in C$^{18}$O emission, exhibiting a 
velocity gradient in the north-south direction on the scale larger than 
2000~AU (SK99). 

One exception in the consistency is the maser feature at \vlsr$\simeq-$7\kms\ that 
was located in the southern part of the southern maser cluster and detected with the 
VLA (SWSC98; SBCSK04) with a big position uncertainty. The maser clusters 
exhibit a velocity gradient in the NNW--SSE direction, but its direction is 
inversed at the vicinity of the peak of continuum emission (SBCSK04). 
Such a complicated velocity distribution cannot be explained by either a pure 
Keplerian rotation disk (figure \ref{fig:model}{\it d}) or a bipolar 
outflow as mentioned above. 

Furthermore, a relative proper motion of two maser features, A and B with (see table 
\ref{tab:pmotions}), in the southern cluster detected in the present work is also consistent with that expected by this model. The vector of the relative motion is roughly consistent with that expected from the model (by $\sim$10\kms\ in the SSE direction), in which the feature B is located much closer to the dynamical center of the disk than the feature A. This relative motion does not appear in a purely Keplerian rotation model (figure \ref{fig:model}{\it d}) because the two maser 
features are located at almost an equal distance from the dynamical center and the two proper motion vectors are also almost equal. The proposed model is also supported by a mean bulk motions of the two maser features with respect to the northern maser cluster mentioned in sect.\ \ref{sec:rmotion}.

Note that the dynamical center of the modeled disk proposed in the present work is 
located at the {\it north} of the two maser features, $\sim$0\arcsec.2 north from the peaks of 3.6~cm and 7~mm continuum emission SCK01, or closer to a peak of 2.6~mm 
continuum emission (SK99). When the dynamical center is put in the south 
of the two maser features or closer to the peaks of 3.6~cm and 7~mm continuum emission 
as shown in figure \ref{fig:model}{\it c}, an extraordinary fast ($\sim$60\kms) 
bulk motion of the YSO with respect to the northern cluster of \h2o maser 
features appears. However, it is still obscure whether these continuum emission 
peaks indicate a single YSO or a YSO binary. 

\section{Discussion}

\subsection{The physical condition on the infalling-rotating disk in the \g192 
southern YSO}
\label{sec:disk-condition}

On a disk proposed in the present paper, it receives gas clumps that 
are impinging on the disk and exciting \h2o maser emission by the shock. The physical condition that can excite the maser emission is described in detail by \citet{fie97}. 
This paper demonstrates the existence of a disk traced by \h2o maser emission 
in IRAS00318$+$6312 (L1287). On the L1287 disk, a pre-shock gas density, a mass 
of the central YSO, and an outer disk radius are estimated to be $n_{H_{2}}\geq 
10^{4}$~cm$^{-3}$ and 3~$M_{\odot}$, and $r_{\mbox{d}}=$4500~AU, respectively. The \h2o maser emission is excited within 35~AU from the YSO, where an impinging gas clump collides with the disk with an impact 
velocity $V_{\theta}\geq$3.1\kms (see eq. \ref{eq:model3}) to be heated over 400~K. 
A gas density at least equal to that in the L1287 disk is expected for the disk 
in the G192 disk from the detection of C$^{18}$O and NH$_{3}$ emission (SBCSK04). In the same gas density as that in L1287 but a higher central mass of $\sim 8$~$M_{\odot}$ and a smaller disk radius ($r_{\mbox{d}}\leq$1000~AU) for 
G192, a radius of maser excitation is expected to be 420~AU ($\sim$0\arcsec.21). 
This radius roughly covers the whole region of the maser emission detected with 
the VLA (SK99; SBCSK04).

Note that the proposed disk model is very thin but still consistent with a thick disk/torus model (SCK01) because \h2o maser may be excited at the mid-plane of the 
disk/torus with the highest gas density 
($n_{\mbox{H$_{2}$}}=10^{9}-10^{11}$~cm$^{-3}$, \cite{eli89}). 

\subsection{A mass accretion on the circumstellar disk and a massive gas disk 
scenario for massive-star formation}
\label{sec:disk-scenario}

On basis of the proposed disk model for the \h2o  maser kinematics, a rate of 
mass accretion on the circumstellar disk is derived as follows, 
\begin{eqnarray}
\dot{M}_{\mbox{disk}}[M_{\odot}\mbox{yr}^{-1}] & = 
& \int^{r_{d}}_{0}2\pi r v_{\theta}\rho 
dr \nonumber \\
& \simeq & 1.6 \times 10^{-7}[n_{\mbox{H$_{2}$}}/10^{4}\mbox{cm}^{-3}]
[M_{\ast}/8\:M_{\odot}]^{1/2}[r_{\mbox{d}}/1000\mbox{AU}]^{3/2},
\label{accretion-rate}
\end{eqnarray} 

\noindent
where  $n_{\mbox{H$_{2}$}}$ is the gas density of the impinging gas clumps, or equal to a pre-shock gas density of a maser clump. When assuming a pre-shock gas density $n_{\mbox{H$_{2}$}}\geq 10^{4}$~cm$^{-3}$ adopted by \citet{fie97}, the obtained 
accretion 
rate is much smaller than a mass loss and infall rates previously estimated, 
5.6$\times$10$^{-4}$~$M_{\odot}$~yr$^{-1}$ and 
$\sim10^{-5}$~$M_{\odot}$~yr$^{-1}$, 
respectively (SK99; SCK01). Note that this disk is an inner part of the region with 
C$^{18}$O emission tracing 
dense gas $n_{\mbox{$H_{2}$}}\geq$10$^{-4}$~cm$^{-3}$ (SK99) and that maser 
excitation models request a much higher pre-shock gas density 
($n_{H_{2}}\sim10^{7}$~cm$^{-3}$, \cite{eli89}). Adopting this density, we obtained 
a accretion rate of $\sim$1.6$\times$10$^{-4}\: M_{\odot}$~yr$^{-1}$. To obtain 
the accretion rate equal to that previously estimated, a higher density, 
$n_{H_{2}}\geq 6\times10^{5}$~cm$^{-3}$ is expected. 

If an estimated accretion rate has persisted since the earliest phase of the star 
formation in G192, total accreted masses for above two accretion rates give 
10~$M_{\odot}$ and 0.6~$M_{\odot}$, respectively. The upper estimated value 
supports a recent conclusion that a massive gas disk scenario for forming a 
massive star is applicable to a stellar object as massive as $\sim$7~$M_{\odot}$ 
\citep{sak05,jia05}. It is also possible to increase in the mass accretion rate at the younger age and to lose a stellar mass due to the outflow. Therefore, it is still difficult for such stellar mass estimation in order to flag proposed scenarios of 
massive star formation performed in G192. 

Note that a scenario in which multiple YSOs merge into a massive star is not completely 
ruled out yet in G192. The existence of a disk with active accretion and a developed  outflow in G192 indicates that these have been continuously developed since their  birth without their disruption by any merger event \citep{rod05}. However, SCK01 and  the present work suggest the existence of more than one YSO near the disk center. If these stars are  gravitationally bound, they might merge in the near future. Such 
a merger scenario  may applicable to a YSO with a mass equal to or larger than those discussed above ($M\geq$10~$M_{\odot}$). 

\subsection{Evolutionary stages of two outflows and two YSOs in \g192}
\label{seq:evolution}

Because the northern and southern YSOs are located within 1600~AU and have 
systemic radial velocities equal within 1\kms (\vlsr$\simeq$6\kms), they 
should be associated with the same parental molecular cloud. SCWC98, SK99, 
SCK01, and SBCSK04 have demonstrated that the southern YSO is a B2-type 
star while the northern YSO is a lower mass star on the basis of a kinematic 
energy of the $^{12}$CO emission outflow and a luminosity of each star  
estimated from the continuum emission, which are correlated each other.  
Properties of the flux density variation and the spatio-kinematics of 
\h2o masers also may indicate some properties or the evolutionary status 
of their driving stellar objects. \h2o masers associated with MYSOs tend 
to have less-collimated spatio-kinematical structures (\cite{ima00, tor01, 
tor03, mos05, god04, god06}, and references therein), while those associated 
with some lower-mass YSOs at the very early stage have highly-collimated 
ones. (\cite{cla98,fur00,set02, fur05}, and references therein). A luminosity 
of \h2o masers associated with an MYSO seems to be higher and temporally more 
stable than that associated with a low-mass YSO. Note that \h2o masers in W75N 
exhibit quite different spatio-kinematics although the radio continuum emission 
properties of their driving stellar objects are quite similar \citep{tor03}. 
It implies that the maser spatio-kinematical structure also evolves as the 
accompanying YSO and its outflow evolve within a period ($t\lesssim10^{4}$~yrs) 
shorter than duration of \h2o maser emission in the same source 
($t\lesssim 10^{5}$~yrs, \cite{gen77}). 

The spatio-kinematical structure of the \h2o masers associated with the 
northern YSO in G192 is higly-collimated as described in \S 3. A luminosity 
of these masers is higher and temporally more stable than that in the southern 
YSO in G192. However our unpublished single-dish spectral date of the \h2o 
masers in G192 shows rapid and significant time variation in the maser 
luminosity as shown in the masers associated with low-mass YSOs. Therefore 
we also suggest that the northern YSO in G192 is a lower mass than the 
southern YSO (B2 star). On the other hand, the \h2o masers associated with the 
southern YSO is tracing an infalling-rotating thin disk, perfectly different 
from those associated with the northern YSO. The southern YSO may be different 
in the stellar mass and the evolutionary status from the northern YSO. 

\citet{she05} (see also references therein) discusses the evolution of 
outflow morphology with the evolution of a massive YSO. In the evolutionary 
scenarios, two sequences of the massive star evolution are considered, 
in the first from a high-mass protostellar object (HMPO) via a hyper-compact 
(HC) H{\rm II} region to a ultra-compact (UC) H{\rm II} region, and in the 
second an O-type star evolves from a B-type star then reaches its final mass 
and stellar luminosity. In both of the sequences, an outflow gets less 
collimated with its evolution. Duration of above evolution may be roughly  
10$^{4}$~yrs \citep{she05}, consistent with that expected from the difference 
in maser spatio-kinematics proposed by \citet{tor03}. In G192, the southern 
YSO is expected to more evolve because of the existence of a more developed 
H{\rm II} region (SBCSK04) and the absence of high collimation of the outflow. 
Note that mass accretion should still persist in the southern YSO through a gas 
disk that may be thinner than that possibly existing around the northern YSO. However, 
7~mm continuum emission is detected in only southern YSO (SBCYK04), 
suggesting enrichment of gas and dust around the southern YSO and the earlier 
stage of the YSO. Taking into account the difference in stellar mass and 
evolution speed, it is concluded that within the same dynamical age, the 
southern YSO with a higher mass evolves more rapidly and its highly-collimated 
jet with \h2o maser excitation disappear sooner than the northern YSO.  
The clearer conclusion of the evolutionary status of these YSOs requests 
information of the gas/dust distribution and the jet morphology in 
higher angular resolution ($\leq$0\arcsec.1).

\section{Conclusions}

We have obtained data of the three-dimensional kinematics of \h2o  masers in 
G192 with short-term ($\sim$2~months) VLBI monitoring observations, which 
were made with the JVN during 2005 March--June, in different epochs from 
those previously 
made with the VLA and the VLBA (SBCSK04). The \h2o masers are associated with 
two MYSOs, which have a highly collimated jet and an infalling-rotating gas disk 
and located at the northern and southern parts in G192, respectively. We have obtained stronger evidence than obtained by previous observations for the existence of an 
infalling-rotating disk with a radius of $\sim$1000~AU and a central 
stellar mass of 5--10~$M_{\odot}$. The absence of any molecular outflow along 
the maser alignment found in the previous VLA observation (SK99) may reject 
the possibility that there exists a jet or an equatorial flow perpendicular 
to the observed molecular outflow with a total length of 5.7~pc. On the basis of the 
spatio-kinematical structure of the jet traced by \h2o maser and $^{12}$CO 
emission as well as assumption of the same dynamical age of the jet and the disk, 
we estimated a dynamical age of the jet/disk system in G192 to be about 
$5.6\times10^{4}$~yrs. With an assumed gas clump density appropriate for maser 
excitation, we have speculated that a massive YSO with a mass of 
$\sim$8~$M_{\odot}$ has an accretion disk that has provided the material onto 
the YSO with a rate of 10$^{-7}-10^{-5}\: M_{\odot}$yr$^{-1}$. These partially 
supports a massive accretion scenario for massive star formation. However, 
there still exist possibility that this massive YSO will become a 
higher-mass star by merging with a lower-mass YSO. 
Observations so as to more clearly elucidate the spatio-kinematics of the \h2o\ 
masers and to directly estimate the pre-shock gas density are crucial for 
estimating a total mass of the existing massive YSO and destiny of the object. 
They also should provide clues for elucidating a history of mass accretion 
and outflow. 

\bigskip
We acknowledge all staff members and students who have helped in array 
operation and in data correlation of the JVN/VERA. H.~I.\ and T.~H.\ 
were supported by Grant-in-Aid for Scientific Research from 
Japan Society for Promotion Science (16540224).

\clearpage

\clearpage

\clearpage

\clearpage 
\setlength{\baselineskip}{6ex}

\begin{table*}[ht]
\input{table1.tex}

\end{table*}

\begin{table*}[h]
\caption{Parameters of the \h2o  maser features identified by 
proper motion toward \g192.} \label{tab:pmotions}
\input{table2.tex}

\end{table*}

\begin{table*}[h]
\caption{Diagonalization analysis of the variance-covariance matrix of 
the velocity and position vectors of the \g192 \h2o  masers.}
\label{tab:vvcm}
\input{table3.tex}

\end{table*}
\end{document}

%% file: table1.tex
\caption{Status of the telescopes, data reduction, and resulting performances in 
the individual epochs of the JVN observations.}\label{tab:status}
\begin{center}
\scriptsize
\begin{tabular}{lccccccc} \hline \hline
& Epoch in & & & Reference & 1-$\sigma$ level & Synthesized & Number of \\
Observation & the year & Duration & Used    
& velocity\footnotemark[2] & noise & beam\footnotemark[3] & detected \\
code & 2005 & (hr) & telescopes\footnotemark[1]  
& (\kms) & (mJy beam$^{-1}$) & (mas)
& features \\ \hline
r05084a \dotfill & March 25 & 10 & MZ, IR, OG, IS, KS, NB\footnotemark[4] 
& 15.8 & 40 & 4.6$\times$0.9, $-$35$^{\circ}$ & 10 \\
r05117a \dotfill & April 27 & 10 & MZ, IR, OG, IS\footnotemark[5], KS, NB 
& 15.6 & 26 & 1.7$\times$1.0, $-$44$^{\circ}$ & 13 \\
r05152a \dotfill & June 1 & 8 & MZ, OG\footnotemark[5], IS\footnotemark[5], KS, NB 
& $-$0.8 & 60 & 1.6$\times$0.9, $-$48$^{\circ}$ & 7 \\ \hline
\end{tabular}

\end{center}
\footnotemark[1] Telescopes that were effectively operated and whose recorded data were valid: 
MZ: the VERA 20~m telescope at Mizusawa, IR: the VERA 20~m telescope at Iriki, OG: the VERA 20~m telescope at Ogasawara Is., IS: the VERA 20~m telescope at Ishigakijima Is., KS: the NiCT 34-m telescope at Kashima, NB: the NRO 45-m telescope at Nobeyama. \\
\footnotemark[2] Velocity channel used for the phase reference in data 
reduction. \\
\footnotemark[3] The synthesized beam made in natural weight; 
major and minor axis lengths and position angle. \\
\footnotemark[4] Ceasing operation for 7 hrs due to strong winds. \\
\footnotemark[5] High system temperature ($>$300~K) due to bad weather conditions.

%% file: table2.tex
\scriptsize
\begin{tabular}{lrrrrrrrr   rrr} \hline \hline    
 Maser feature\footnotemark[1]
 & \multicolumn{2}{c}{Offset\footnotemark[2]}
 & \multicolumn{4}{c}{Proper motion\footnotemark[2]}
 & \multicolumn{2}{c}{Radial motion\footnotemark[3]}
 & \multicolumn{ 3}{c}{Peak intensity at three epochs} \\                           
 (G192.16$-$3.84: 
 & \multicolumn{2}{c}{(mas)} 
 & \multicolumn{4}{c}{(mas yr$^{-1}$)}
 & \multicolumn{2}{c}{(km s$^{-1}$)}
 & \multicolumn{ 3}{c}{(Jy beam$^{-1}$)} \\                                      
 & \multicolumn{2}{c}{\ \hrulefill \ } 
 & \multicolumn{4}{c}{\ \hrulefill \ } 
 & \multicolumn{2}{c}{\ \hrulefill \ } 
 & \multicolumn{ 3}{c}{\ \hrulefill \ } \\                                       
 I2006) & R.A. & decl. & $\mu_{x}$ & $\sigma \mu_{x}$ & $\mu_{y}$ 
 & $\sigma \mu_{y}$ & V$_{z}$ & $\Delta$V$_{z}$\footnotemark[4]
 & Epoch  1& Epoch  2& Epoch  3 \\ \hline                                        
  1\footnotemark[5]   \ \dotfill \  &$     2.90$&$   -41.46$&$  -0.78$&   0.39 
&$  -1.03$&   0.20 &$  -6.76$&   1.05
  &        1.93 &        2.44 &        5.87   \\                                 
  2   \ \dotfill \  &$  -258.37$&$   479.06$&$  -0.34$&   0.58 &$  -2.26$&   0.24
 &$  -3.59$&   0.49
  &        1.63 &        1.70 &        0.99   \\                                 
  3   \ \dotfill \  &$  -259.06$&$   479.24$&$   0.23$&   2.38 &$  -2.64$&   0.67
 &$  -2.96$&   0.52
  &        1.32 &        0.83 &    ...        \\                                 
  4   \ \dotfill \  &$  -246.15$&$   500.19$&$  -4.56$&   0.52 &$  -7.15$&   0.38
 &$  -2.76$&   0.84
  &        2.18 & ...\footnotemark[6] &        0.61   \\                                 
  5   \ \dotfill \  &$  -260.82$&$   479.38$&$   0.93$&   0.86 &$  -3.12$&   0.13
 &$  -2.65$&   0.84
  &        3.80 &        3.75 &        1.99   \\                                 
  6   \ \dotfill \  &$  -246.24$&$   500.08$&$  -6.39$&   0.26 &$  -7.45$&   0.21
 &$  -1.73$&   2.32
  &        5.10 &        5.85 &       47.60   \\                                 
  7   \ \dotfill \  &$  -246.61$&$   499.88$&$  -8.80$&   1.20 &$  -8.87$&   0.95
 &$  -0.85$&   0.47
  &        0.38 &        1.18 &    ...        \\                                 
  8   \ \dotfill \  &$    46.63$&$   657.13$&$   3.74$&   0.79 &$  -2.99$&   0.79
 &$   9.86$&   0.47
  &    ...      &        0.88 &        0.69   \\                                 
  9\footnotemark[7]   \ \dotfill \  &$     0.00$&$     0.00$&$   0.00$&   0.13 
&$   0.00$&   0.47 &$  15.87$&   1.76
  &        9.19 &       12.29 &        7.59   \\                                 
 \hline
 \end{tabular}

\noindent
\footnotemark[1]
Water-maser features detected toward G192.16$-$3.84. The feature is designated as 
G192.16$-$3.84:I2006 {\it N}, where {\it N} is the ordinal source number given in this 
column (I2006 stands for sources found by Imai et~al. and listed in 2006). \\
\footnotemark[2]
Relative value with respect to the motion of the position-reference maser feature: 
G192.16$-$3.84:I2006 {\it 9}. \\
\footnotemark[3] Relative value with respect to the local stand of rest. \\
\footnotemark[4] Mean full velocity width of a maser feature at half intensity. \\
\footnotemark[5] The maser feature A (see main text). \\
\footnotemark[6] The maser feature was spatially overlaped with the maser feature 
G192.16$-$3.84:I2006 {\it 6} and its peak intensity was unable to evaluated. \\
\footnotemark[7] The maser feature B (see main text). \\

%% file: table3.tex
\begin{tabular}{ccc} \hline \hline
& \multicolumn{2}{c}{Egenvector} \\
Egenvalue & Inclination & Position angle \\ 
$ \left[(\mbox{km~$^{-1}$})^{2} \right]$ & & \\ \hline
\multicolumn{3}{c}{Velocity variance-covariance matrix} \\ \hline
$2502\pm306$ & 2\arcdeg.0$\pm$0\arcdeg.5 & 59\arcdeg.6$\pm$2\arcdeg.9 \\
$106\pm49$ & $-$20\arcdeg.0$\pm$7\arcdeg.8 & $-$29.3$\pm$8\arcdeg.7 \\ \hline
\multicolumn{3}{c}{Spatial variance-covariance matrix} \\ 
(mas$^{2}$) & & \\ \hline
16939 & ... & 60\arcdeg.6 \\ \hline
\end{tabular}

%% file: PASJ-WB724_060830.bbl
\clearpage
\begin{thebibliography}{99}
\bibitem[Bally \&  Zinnecker(2005)]{bal05}
Bally,~J., \&  Zinnecker,~H.\ 2005, \aj, 129, 2281
\bibitem[Bloemhof(2000)]{blo00}
Bloemhof,~E.~E.\ 2000, \apj, 533, 893
\bibitem[Bonnell \& Bate(2006)]{bon06}
Bonnell,~I.~A., \& Bate,~M.~R.\ 2006, \mnras, in press (astro-ph/0604615)
\bibitem[Bonnell, Vine, \& Bate(2004)]{bon04}
Bonnell,~I.~A.,Vine,~S.~G., \& Bate,~M.~R.\ 2004, \mnras, 349, 735
\bibitem[Cassen \& Moosman(1981)]{cas81}
Cassen,~P., \& Moosman,~A.\  1981, Icarus, 48, 363
\bibitem[Chini \etal(2004)]{chi04}
Chini~R., Hoffmeister,~V., Kimeswenger,~S., Nielbock,~M., N\"{u}rnbrgue,~D., 
Schmidtobreick,~L., \& Sterzik,~M.\ 2004, Nature, 429, 155
\bibitem[Chikada \etal(1991)]{chi91}
Chikada,~Y., Kawaguchi,~N., Inoue,~M., Morimoto,~M., Kobayashi,~H., 
Mattori,~S., Nishiura,~T., Hirabayashi,~H.\ \etal\ 1991, in Frontiers of VLBI, 
ed.\ H.~Hirabayashi, M.~Inoue, \&  H.~Kobayashi 
(Tokyo: Universal Academy Press), p.~79
\bibitem[Claussen \etal (1998)]{cla98}
Claussen,~M.~J., Marvel,~K.~B., Wootten,~A., \&  Wilking,~B.~A.\  1998, 
\apj, 507, L79
\bibitem[Devine \etal (1999)]{dev99}
Devine,~D., Bally,~J., Reipurth,~B., Shepherd,~D.~S., \&  Watson, ~A.~M.\ 
1999, \aj, 117, 2919
\bibitem[Diamond(1995)]{dia95}
Diamond,~P.~J.\  1995, in ASP Conf.~Ser.~82, VERY LONG BASELINE 
INTERFEROMETRY AND THE VLBA, ed.\ J.~A.~Zensus, P.~J.~Diamond, 
\&  P.~J.~Napier (San Francisco: ASP), p.~227
\bibitem[Elitzur, Hollenbach, \&  McKee(1989)]{eli89}
Elitzur,~M., Hollenbach,~D.~J., \& McKee,~C.~F.\ 1989,  \apj, 346, 983
\bibitem[Fiebig(1997)]{fie97} Fiebig,~D.\  1997, \aap, 327, 758 (F97)
\bibitem[Furuya\etal(2005)]{fur05}
Furuya,~R.~S., Kitamura,~Y., Wootten,~H.~A., Claussen,~M.~J., \& Kawabe,~R.\  
2005, \aap, 438, 571
\bibitem[Furuya\etal(2000)]{fur00}
Furuya,~R.~S., Kitamura,~Y., Wootten,~H.~A., Claussen,~M.~J., Saito,~M., 
Marvel,~K.~B., \& Kawabe,~R.\  2000, \apj, 542, L135 
\bibitem[Genzel \&  Downes(1977)]{gen77}
Genzel,~R., \&  Downes,~D.\  1977, \aaps, 30, 145
\bibitem[Goddi \& Moscadelli(2006)]{god06}
Goddi,~C., \&  Moscadelli,~L., 2006, \aap,  447, 587
\bibitem[Goddi\etal(2004)]{god04}
Goddi,~C., Moscadelli,~L., Alef,~W., \&  Brand,~J.\ 2004, \aap, 420, 929
\bibitem[G\'omez \etal(2005)]{gom05}
G\'omez,~L., Rodr\'iguez,~L.~F., Loinard,~L., \&  Lizano,~S.\ 2005, \apj. 
635, 1166
\bibitem[Honma \etal(2003)]{hon03}
Honma,~M.,Fujii,~T., Hirota,~T., Horiai,~K., Iwadate,~K., Jike,~T., Kameya,~O., 
Kamohara,~R., \etal\  2003, \pasj, 55, L57
\bibitem[Imai, Iwata, \& Miyoshi 1999]{ima99}
Imai,~H., Iwata,~T., \&  Miyoshi,~M.\   1999, \pasj, 51, 473
\bibitem[Imai \etal(2000)]{ima00}
Imai,~H., Kameya,~O., Sasao,~T., Miyoshi,~M., Deguchi,~S., Horiuchi,~S., \&  
Asaki,~Y.\  2000, \apj, 538, 751
\bibitem[Indebetouw \etal(2003)]{ind03}
Indebetouw,~R., Watson,~C., Johnson,~K.~E., Whintney,~B., \&  Churchwell,~E.\ 2003,
\apj 596, L83
\bibitem[Jiang\etal(2005)]{jia05}
Jiang,~Z., Tamura,~M., Fukagawa,~M., Hough,~J., Lucas,~P., Suto,~H., Ishii,~M., \& 
Yang,~J.\ 2005, Nature, 437, 112
\bibitem[Keto \& Wood(2006)]{ket06}
Keto,~E., \&  Wood,~K.\ 2006, \apj, 637, 850
\bibitem[Keto(2003)]{ket03}
Keto,~E.\ 2003, \apj, 599, 1196
\bibitem[Krumholtz, McKee, \&  Klein(2005)]{kru05}
Krumholtz,~M.~R., McKee,~C.~F., \&  Klein,~R.~I.\ 2005, \apj, 618, L33
\bibitem[Moscadelli, Cesaroni, \&  Rioja(2005)]{mos05}
Moscadelli,~L., Cesaroni,~R., \&  Rioja,~M.~J.\ 2005, \aap, 438, 889
\bibitem[Nagano(1989)]{nag89}
Nagano,~T.\ 1989, \apj, 345, 464
\bibitem[Reid \&  Moran(1981)]{rei81} 
Reid,~M.~J., \&  Moran,~J.~M.\  1981, \araa, 19, 231
\bibitem[Rodr\'iguez\etal(2005)]{rod05}
Rodr\'iguez,~L.~F., Poveda,~A., Lizano,~S., \&  Allen,~C.\ 2005, \apj, 627, L65
\bibitem[Sako\etal(2005)]{sak05}
Sako,~S., Yamashita,~T., Kataza,~H., Miyata,~T., Okamoto,~Y.~K., Honda,~M., 
Fujiyoshi,~T., \&  Terada,~H., \etal.\ 2005, Nature, 434, 995 
\bibitem[Seth\etal(2002)]{set02}
Seth,~A.~C., Greenhill,~L.~J., \&  Holder,~B.~P.\ 2002, \apj, 581, 325
\bibitem[Shepherd(2005)]{she05}
Shepherd,~D.\ 2005, in IAU symposium 227, eds. R.~Cesaroni, M.~Felli, 
E.~Churchwell \& M.~Walmsley (Cambridge: Cambridge Univ. Press), p.~237
\bibitem[Shepherd\etal(2004)]{she04}
Shepherd,~D.~S., Borders,~T., Claussen,~M.~J., Shirley,~Y.\ \&  Kurtz,~S.~E.\ 2004, \apj, 614, 211 (SBCSK04)
\bibitem[Shepherd, Claussen, \&  Kurtz(2001)]{she01}
Shepherd,~D.~S., Claussen,~M.~J., \& Kurtz,~S.~E.\ 2001, Science, 292, 1513 
(SCK01)
\bibitem[Shepherd, \&  Kurtz(1999)]{she99}
Shepherd,~D.~S., \& Kurtz,~S.~E.\ 1999, \apj, 523, 690 (SK99)
\bibitem[Shepherd\etal(1998)]{she98}
Shepherd,~D.~S., Watson,~A.~M., Sargent,~A.~I., \&  Churchwell,~E.\ 1998, 
\apj, 507, 861 (SWSC98)
\bibitem[Snell\etal(1990)]{sne90}
Snell,~R.~L., Dickman,~R.~L., \&  Huang,Y.-L.\ 1990, \apj, 352, 139
\bibitem[Torrelles \etal (2003)]{tor03}
Torrelles,~J.~M., Patel,~N.~A., Anglada,~G., G\'omez,~J.~F., Ho,~P.~T.~P., Lara,~L., Alberdi,~A., Cant\'o,~J., \etal\  2003, \apj, 598, L115
\bibitem[Torrelles \etal (2001)]{tor01}
Torrelles,~J.~M., Patel,~N.~A., G\'omez,~J.~F., Ho,~P.~T.~P., Rodor\'iguez,~L.~F., 
Anglada,~G., Garay,~G., \& Greenhill,~L.~J., \etal\  2001, \apj, 560, 853
\end{thebibliography}
